# The 2017 Merrimack College Mobile Lightboard: Personalized Office-Based Lightboarding on a Shared Mobile Board

Craig W. Looney[1,2], Christopher L. Duston[1]
James H. Biegel[1], Manuel A. Carrero Jr.[1], Nicholas A. Valente[1]

[1] *Merrimack College, 315 Turnpike Street, North Andover, MA 01845*
[2] *Corresponding author: looneyc@merrimack.edu*

**ABSTRACT.** We present and evaluate the Merrimack College Mobile Lightboard, which we built in 2017 and which is still in use in 2025. Our design was closely based on that of the groundbreaking 2014 Duke University Lightboard but with a smaller glass pane, 5 feet wide instead of 6 feet wide. The resulting scaled-down board is small enough to move in and out of individual faculty offices, but large enough to make detailed videos. An updated construction cost estimate is provided, and recent upgrades are briefly discussed. We believe that personalized office-based lightboarding represents an attractive option for many faculty members, with strengths that complement those of centralized installations. Our successful build suggests that this option is technically and financially feasible at many if not most higher education institutions.

## Introduction

The lightboard – independently invented by Michael Peshkin [1] and Matthew Anderson [2] – is a device for producing high-quality instructional videos, consisting of a transparent pane rimmed with inward-directed LED lights. The presenter writes on the glass with fluorescent markers and speaks while facing the video camera through the glass; the marker residue disrupts the total internal reflection that would otherwise occur (frustrated total internal reflection), allowing light to exit the glass through and illuminate the fluorescent marker writing [1].[1] While centralized installations rapidly came to be seen as the lightboarding gold standard, from the very beginning there has also been substantial interest in *personal* lightboards, as evidenced by numerous YouTube videos documenting personal DIY builds. We advocate for a big tent perspective that recognizes the complementary strengths of both centralized and personalized setups for expanding accessibility, enhancing user experience, and driving advancements in lightboard design and technology. For example, centralized studios can provide access to – and provide an essential source of R&D funding for – the ongoing development[2] of high-end devices and features; if staffed with professionals skilled in production and editing, they can substantially reduce technical prerequisites for potentially interested faculty members. On the other hand, personal installations have complementary advantages including but not limited to on-demand access, freedom from one-size-fits-all recording blocks, enhanced privacy, and a greater degree of overall control. In addition, interest in personal lightboarding – and personally-sized lightboards – contributes to the viability and ongoing R&D of commercial lightboard manufacturers, whose personally-sized offerings in 2025 are substantially better and less expensive than in 2017; meanwhile, DIY and institutionally-supported user-builders working within the personalized framework have made, and continue to make, important and distinctive contributions to design, technology, and practice.

---

[1] While edge illumination and frustrated total internal reflection are integral elements of the Peshkin [1] and Anderson [2] designs, and are also normally considered to be a defining characteristic of what most people mean when they say the word "lightboard," it is actually possible to noticeably – if less spectacularly – achieve the effect of illuminated writing against a dark background without edge illumination, as demonstrated in [3].

[2] To clarify: While high-end development, which was largely carried out in-house in the early days, is now largely carried out by commercial manufacturers, the demand for and funding for high end lightboards with high end features – beyond the lowest-common-denominator feature of giving the appearance of writing on air while facing the camera – continues to be largely driven by centralized institutional installations.

The 2017 Merrimack College Mobile Lightboard described in this paper represents a successful early attempt to design and build a shared personal lightboard that can be used privately within a small faculty office. This robust build, which is still in use in 2025, features a T-slotted aluminum frame and a 3.5-foot by 5-foot pane of low-iron glass. It is small enough to roll into offices and stockrooms, but large enough to make videos that present reasonably-detailed introductory physics examples. In undertaking this build, we relied heavily on Chip Bobbert's description [4] and documentation [5] of the influential 2014 Duke University lightboard build. While that board featured a large 4-foot by 6-foot pane of low-iron glass, it was capable of smooth multi-directional movement within a large space. The reduced width and weight of our scaled-down board makes it easier for a single person to control, and makes it easier to move into and out of – and maneuver within – relatively small office spaces or stockrooms.

After presenting an overview of our build, we provide an updated estimate of construction costs and briefly describe more recent upgrades. Build details beyond the overview presented in the text are presented in an appendix following the references.

## 2017 design and build

*Frame overview.* As noted above, our frame design was based closely on that of the 2014 Duke University build [4, 5], and used the same 15 series T-slotted aluminum framing, from the same supplier (80/20). We used the standard framing rather than light or ultra-light framing; however, for shorter pieces, either of the lighter framing options would probably have been fine. To scale down from a 6-foot wide pane to a 5-foot wide pane, we simply reduced the dimensions of all the long transverse T-slotted aluminum pieces by 12 inches. To scale the glass height from 4 feet to 3.5 feet, we made appropriate adjustments to the locations of the long transverse T-slotted aluminum pieces supporting the frame. The overall height was adjusted to ensure that the completed board could pass through all doorways from the stockroom to faculty offices, and be moved into or out of our building's small elevator, so that it could be moved between floors. We also made sure that the board could fit through the building exit and into other important campus spaces. We simplified the vertical support design somewhat by replacing, on each side of the board, a 36-inch long support piece (attached at a custom angle using ball joints in the Duke design) with two shorter 45-degree support pieces. There were a handful of further minor changes. The frame we built offers plenty of stability, so future builders need not be overly concerned that their own further carefully-considered adjustments or simplifications might result in catastrophic failure. We present our build not as something that must be carefully copied, but rather as an example of something that has worked that can inform incremental evolution or inspire radically different designs. Jumping ahead somewhat, Fig. 1 shows two views of the finished board, including the frame.

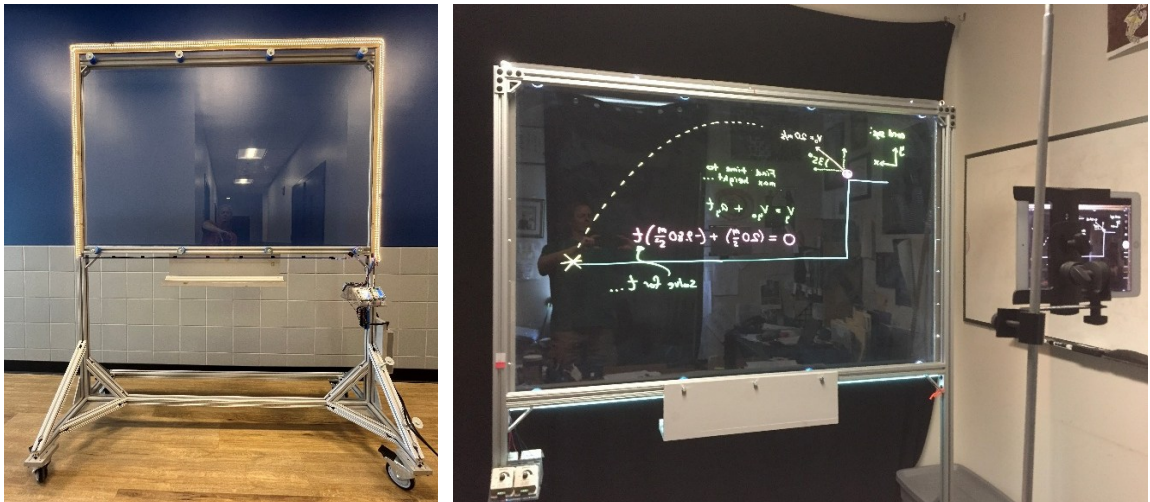

**Figure 1. LEFT:** a full view of the "presenter" side of the finished board. **RIGHT:** a view of the "camera" side of the finished board, in "lightboarding position" in front of a black backdrop in a faculty office.

*Frame details.* In *Appendix F*, which follows after the references, we present the frame parts list, with labeled pictures to indicate where the parts were used. Here, we note that the T-slotted aluminum framing and associated hardware were purchased from 80/20 and we were very satisfied with the quality. One of the co-authors (CWL) has used lighter 15-series T-slotted profiles from a budget supplier (TNutz) in a subsequent lightboard project and noticed no difference in quality or functionality for lightboard framing.

*Low-iron tempered glass and mounting hardware.* We used a 3.5-foot by 5-foot pane of 3/8-inch thick tempered low-iron glass[3] from a local supplier. The supplier referred to the glass as "Starfire" and confirmed the spelling (with an "f") when asked; this differs from the "ph" spelling of the Starphire low-iron glass used in and popularized by the Peshkin build [1]. We did not ask further questions at the time, but it turns out that the word "Starfire" can refer to a somewhat lower grade of low-iron glass (made by the makers of Starphire glass) but also is commonly used as a generic name for low-iron glass. We do not know whether we have generic starfire or brand-name Starfire, and we have no way to reliably find out at this point. Whatever the case may be, the glass has proven to be plenty clear. For example: until recently, we had LEDs mounted on only the top and side edges but not on the bottom edge; while there is a noticeable intensity gradient when the edge lights are dim, it is normally not noticeable at the higher edge light levels used when making videos as shown in Fig. 2.

---

[3] The clarity and transmissibility of low-iron glass are superior to that of ordinary glass. This is important, especially for medium-sized or large boards, to ensure that the intensity of light from edge-mounted LEDs does not diminish appreciably at pane locations far from the edge. The glass is *tempered* for strength and safety. While plexiglass is stronger, safer, and lighter than glass, has lightboard-related optical properties that rival those of low-iron glass, can be easily drilled, and has been successfully used in numerous low-budget builds, it scratches easily. Since we wanted maximal long-term scratch resistance, we chose to use (tempered) low-iron glass rather than plexiglass. We were drawn to the Duke design [4, 5] because of its potential to provide robust mobile support for a large glass pane. A plexiglass pane would have required far less support.

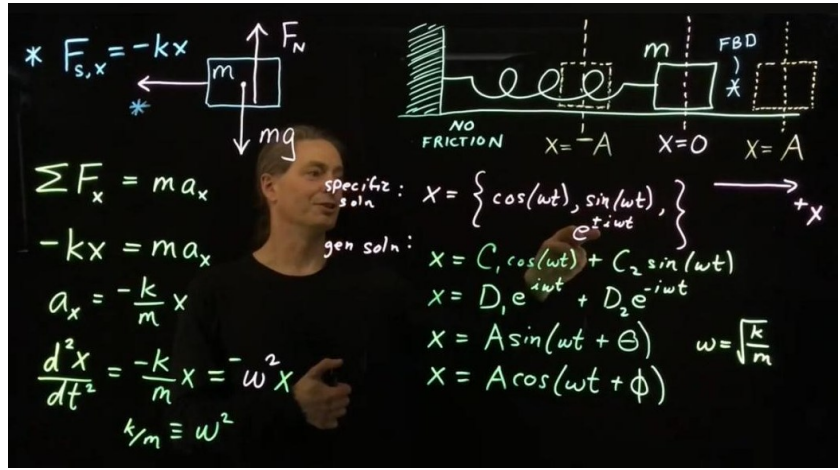

**Figure 2.** Frame from a 2018 video [6]. As noted in the body of the text, the top to bottom brightness gradient is not noticeable, even though the bottom edge of the glass has no edge illumination.

Following the Peshkin [1] and Duke [4, 5] designs, we had the supplier drill mounting holes in the glass prior to tempering (glass cannot be drilled after tempering), so that the glass could be mounted on the frame using a 3D-printed glass standoff system [7] and 5/16-18 bolts (which fit the standard T-nuts of the "15 series" T-slotted aluminum framing). We had originally planned to print the standoffs designed by Chip Bobbert [8] for the Duke build, but we wanted slightly slimmer standoffs, so one of us (NAV) developed a lower profile 3D printable design [7]. The standoffs and bolts are shown in Fig. 3. The as-printed standoffs seemed quite slippery against the glass – much more slippery than finger pads – so we cut thin vinyl washers (also shown in Fig. 3) from a thin flexible vinyl sheet purchased from a fabric store. These grippy washers allow static friction to more easily contribute support without over-tightening the bolts, and also provide a measure of cushion between the hard plastic standoffs and the glass.

The specification sketch we gave to the glass supplier showing the glass dimensions and hole locations is shown in Fig. 4. The distances from the hole centers to the nearest edge are slightly larger than those used in the Duke build. The hole clearances (9/16-inch diameter hole, to fit a "standoff" sleeve with an outer diameter of 1/2 inch) are also slightly larger. Even with these "larger" 1/16-inch hole clearances, and with the 1/16 inch or so of further play allowed by the slot system, a very careful alignment and spacing of the support slots – and pre-positioning of the nuts for securing the standoffs – is essential. Near dead-on accuracy for the hole centers was also essential, but was beyond our control. Fortunately, our assembly and the drilling were both within spec, and the glass installation went smoothly.

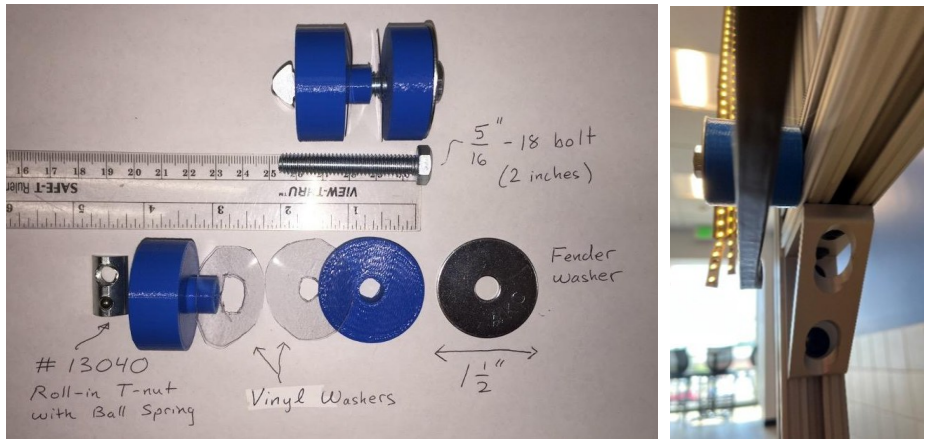

**Figure 3. LEFT:** Photograph of the assembled and disassembled standoff system. See [7] for the 3D print plan for the blue parts. The part number shown for the 15-series roll-in T-nut (for the 5/16-18 bolt) is the number in the 80/20 system. **RIGHT:** Standoff shown in-situ.

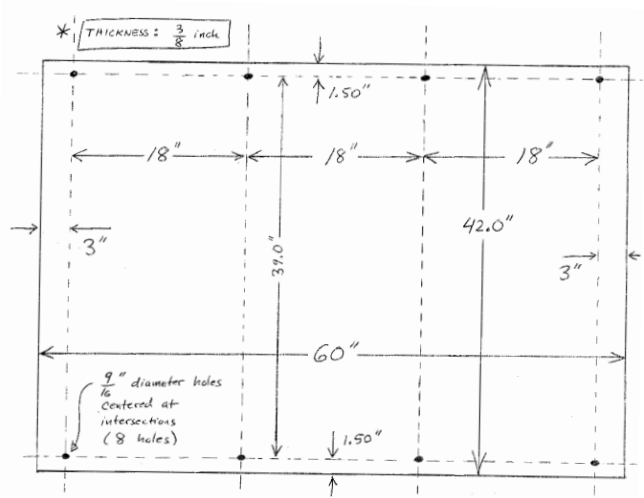

**Figure 4.** Image of actual specifications sketch submitted to the glass supplier showing glass dimensions and hole-center locations, and hole diameters. The glass was drilled prior to tempering.

***LED lighting and mounting clips (copied from the Duke University design).*** Closely following the Duke University design [4, 5], two 3000K LED strips were used for presenter[4] illumination and one 5000K LED strip was used for pane illumination. All three were 12 Volt SMD 2835 LED[5] strips with 120 LEDs per meter. These were mounted in the same configuration as in the

---

[4] CWL has used 6000K LEDs for presenter lighting in subsequent builds. While he thinks he prefers those, he has not undertaken a systematic head-to-head comparison. Also, when he was heavily using the 3000K lighting he was still learning how to optimize camera settings.

[5] SMD 2835 LEDs are more efficient than earlier generation strands (SMD 3528, SMD 5050). They are available in various "temperature" ranges (which impacts how yellowish or bluish the light appears) and in various chip densities up to 240 LED/meter. For white light LEDs used for presenter or glass illumination, CWL – who as of September 2025 has built 3 subsequent lightboards – has found 120 LED/meter strands to be more than sufficient.

Duke build using the same 3D-printed LED holders (3D print file available at [9]) that mount directly onto the top and side edges of the 3/8-inch thick glass edge; these clips have one slot that directs the 5000K pane illumination into to the glass edge, and two slots – at different angles – for concentric 3000K presenter LED strands (Fig. 5; see also the Duke build information [4, 5]). Following the Duke build, until recently we left the bottom glass edge unilluminated.

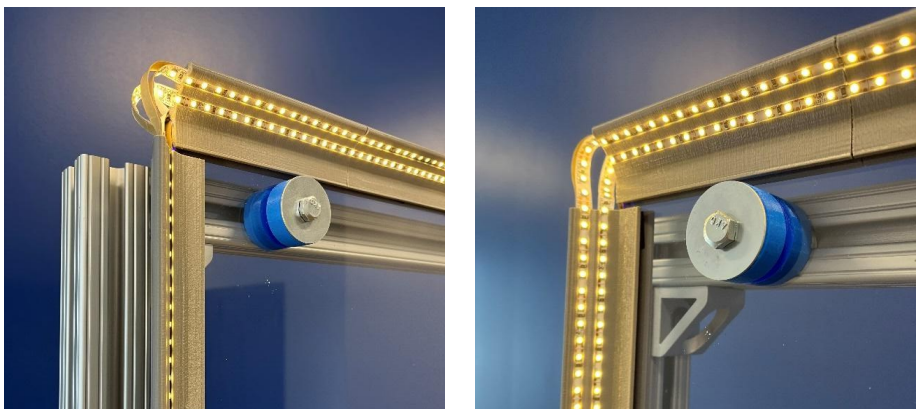

**Figure 5.** Two views of the same corner of the glass pane, showing edge-mounted LED clips with concentric strands of 3000 K presenter LEDs on. A glass standoff is also shown.

*LED power and illumination control.* With some minor modifications, we mostly followed the Duke design [4, 5]: a pair of 150-Watt LED drivers and a terminal block distribute parallel 12 Volt inputs to three PWM dimmers (Fig. 6, LEFT), so that the glass-edge and concentric presenter strands can be separately controlled. In retrospect, a single higher-powered driver would have been simpler. CWL has used a single 150-Watt LED driver in a subsequent build with a larger 4-foot by 6-foot pane to drive all glass and presenter lighting.

While pulse wave modulation (PWM) dimmers like those used in the Duke build – and originally used in our build as noted above – are still predominantly used in 2025 in both independent and commercial builds, the modulation frequencies they introduce into the recording process can produce unsightly banding in recorded video [10]. Perhaps even more importantly, they offer relatively coarse illumination control, and it is virtually impossible to obtain precisely repeatable illumination levels unless the knobs are never turned. Nice step-down DC "buck" converters – which provide a steady but adjustable DC voltage – with digital readouts for both voltage and current (Fig. 6, CENTER) are currently available for less than $20 each;[6] in 2023 one of us (CWL) installed these on his own home-built lightboard and on the Merrimack College Mobile Lightboard. Using these in place of PWM dimmers eliminates the possibility of banding and allows the fine, repeatable control necessary for careful optimization [10]. Another alternative is to use a separate 12 Volt variable DC power supply with a digital readout (Fig. 6,

---

[6] The one drawback of DC buck converters is that they normally reduce the maximum voltage. The nice buck converters described here reduce the maximum voltage from 12 Volts to about 11.2-11.3 Volts. Less expensive buck converters may reduce the maximum output voltage to well below 11 Volts. Bargain buck converters also tend to have less useful readouts and may require a small jewelers screwdriver to adjust the output voltage. The expected reduction in maximum output voltage should be noted in the device specifications.

RIGHT) to control each strand; while these do not offer the same degree of finely repeatable control as the just-described buck converters, the digital readout allows more repeatable control than PWM dimmers and they don't contribute to banding.

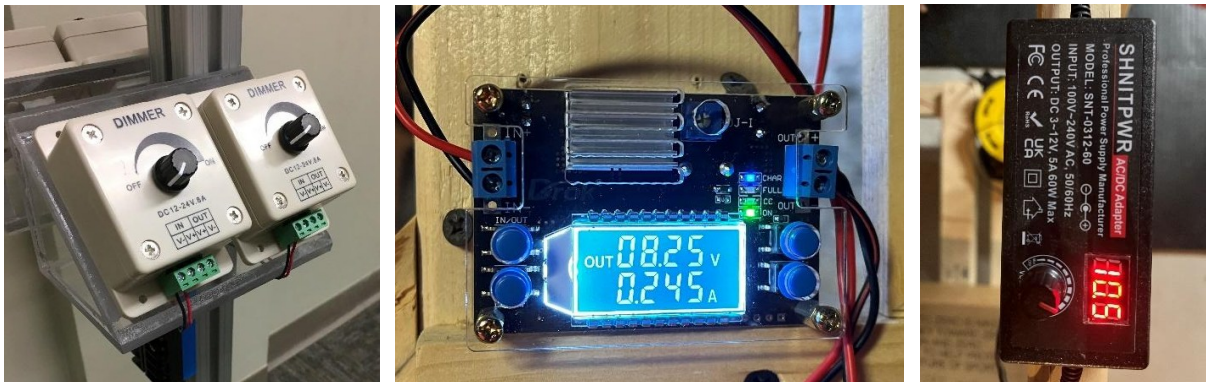

**Figure 6. LEFT:** The original PWM dimmers. **CENTER:** DC "buck" converter with digital voltage and current readouts. Buttons allow voltage to be stepped up or down in 0.05 Volt increments, allowing fine illumination control and repeatability. **RIGHT:** Variable DC power supply with digital readout. A single turn knob allows repeatability to about 0.1 Volts.

*Further miscellaneous build notes.* We used the same commercial terminal block and the same 3D printed mount as in the Duke build [4, 5]. The 3D printed mount was great, but the terminal block required modification. Rather than recommend a specific make and model of terminal block, we suggest simply searching online sellers for "terminal blocks" and choosing one – or more – that look attractive. We prefer to use multiple inexpensive terminal blocks to avoid cramped connections. Those who use variable 0-12 Volt DC power supplies may not need terminal blocks at all, since the variable DC power supplies plug directly into 120-Volt outlets.

We did not use the Duke 3D print plans for the 150-Watt power supply mounts or for the PWM dimmer mounts. The 150-Watt power supplies were simply mounted with screws onto a ¾-inch thick piece of wood. The wood was bolted to the frame using 5/16-18 bolts, which fit the standard 15-series T-nuts (Fig. 7, LEFT). This scheme – using ¾-inch thick wood bolted to the T-nuts in the 15-series frame slots using 5/16-18 bolts – is a fast and easy way to attach just about anything that involves small screws to a 15-series T-slotted aluminum frame. More generally, the slot system allows all manner of custom-fabricated or home-built elements – such as the home-built wooden marker tray shown below (Fig. 7, RIGHT) or the custom-fabricated[7] plexiglass dimmer mounts (shown previously above in Fig. 6, LEFT) – to be attached to the frame using 5/16-18 bolts.

*Brief overview of video setup and production process.* To carry out office-based video recording, CWL and CLD installed blackout blinds in their offices, which can be fully closed to block all external light that might directly reflect back into the camera, or which might illuminate objects on the camera side of the pane to create visible reflections. Black fabric backdrops, like the one shown in Fig. 1 (RIGHT), are temporarily hung during recording. From the very beginning, we have used our college-issued iPads or our mobile phones to record video, rather than the dedicated video cameras recommended by other earlier builders. As emphasized by Steve Griffiths [11], attention to camera settings is important. For mobile devices with fixed apertures, the ISO and shutter speed settings are the only means to adjust exposure and light sensitivity. Consequently CWL [10] strongly recommends – for lightboarders who use mobile devices to record video – the use of 3$^{rd}$ party video apps that allow manual control of ISO, shutter speed, white balance, and frame rate.[8] For video editing, we use apps or programs that are free or for which we have access through institutional subscriptions, and we normally make our finished videos publicly available on our respective YouTube channels.[9]

---

[7] Credit: Douglas Therrien, Laboratory Director/Technician, Merrimack College School of Engineering and Computational Sciences.

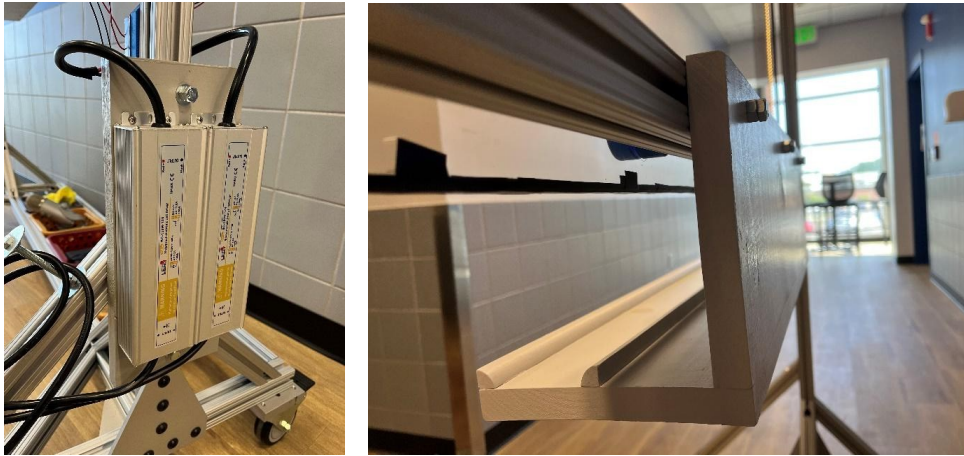

**Figure 7. LEFT:** The twin 150-Watt power supplies were screwed onto a ¾-inch thick board that was mounted to the frame using 5/16-18 bolts. **RIGHT:** A home-built marker tray made from ¾-inch thick wood, bolted directly to the frame using 5/16-18 bolts.

***Brief remarks on recent upgrades.*** As noted above, in 2023 the PWM dimmers were replaced with DC step-down converters, which eliminate the possibility of banding and which offer much better and much more repeatable illumination control. In 2024 we installed near-visible ~395 nm UV blacklight LEDS (SMD 2835 with 120 LEDs per meter). To do this, we simply pulled off the clips – without removing any of the LEDs from the clips – and taped the blacklight LEDs directly to the glass edge, and then we put the clips back on over the top of the edge-taped blacklight LEDs. This will allow us to continue to use the original 3000K presenter LEDs. This was motivated by initial results presented in 2024 by CWL [12] that suggest ~395 nm blacklight pane illumination has advantages for masking smudges and glass flaws; we note that McCorkle and Whitener reported a qualitative preference for lightboard writing illuminated by blacklight [13, 14]. While the motivation for blacklight use is sound and initial post-upgrade use has been reasonably positive, further discussion of blacklight is beyond the scope of this particular paper.

---

[8] Settings are camera specific and also are impacted by the details of pane and presenter lighting. For any given installation there are many combinations of camera settings and lighting levels that will give an attractive result, and many more that will not. Consequently, obtaining an attractive result involves iterative adjustment. Steve Griffiths's video [11] offers useful guidance.

[9] https://www.youtube.com/@levitopher and https://www.youtube.com/@CraigWLooney.

*A brief note on markers and erasing.* We have found that Expo Neon and Quartet Neon markers glow attractively and are easily erasable with a dry microfiber cloth. We have found that orange and red Crafty Croc markers look great when using ~395 nm blacklight pane illumination; the Crafty Croc orange is brighter than the Expo Neon orange (Quartet has no orange), and the Crafty Croc red – in addition to being the brightest red we have found – has a distinctly different shade than the Expo and Quartet reds under blacklight. However, the Crafty Croc markers take a bit longer to erase, although they are still fully erasable using a dry microfiber cloth. With appropriate attention to lighting and camera settings – as Steve Griffiths advises in [11] – carefully wiping with a dry microfiber cloth should be all that is normally required between uses to erase the board and render smudges effectively invisible.

## Build costs

The cost of all materials for our 2017 build – T-slotted aluminum framing and hardware from 80/20, custom-drilled low iron "starfire" glass (whatever that was), LED lights and associated electrical components – was about $2500. The cost would be substantially higher in 2025, likely between $3500 and $4000. However, there are reasonable avenues available to substantially reduce costs. Low-iron glass can now be purchased mail-order, with free shipping, from several suppliers.[10] T-slotted aluminum can be purchased from economical suppliers such as TNutz. Economical high-quality casters are widely available (there is no need to spend $35 per caster like we did in 2017) and wood blocks can be used in place of expensive caster mounts. Less expensive "Lite" or "Ultra-Lite"[11] (vs. the more expensive "standard") framing can be used throughout; double-width profiles can be used where extra support is needed. Based on a recent build, CWL estimates that a high-quality aluminum-framed mobile lightboard with a 3.5-foot by 5-foot pane of mail-order low-iron glass could be built for $2500 in 2025.

DIY builders capable of wood-based construction can easily build wood-based mobile lightboard frames. While a mobile wood-based frame capable of safely supporting a 5-foot wide glass pane might be a bit bulky, the mobile home lightboard CWL built in 2020 [10] with a 4-foot wide mail-order low-iron glass pane would be sufficiently mobile to use as a shared office-based lightboard. This was built for just over $800 in 2020.

---

[10] Since online suppliers do not normally have the capacity to drill custom mounting holes, appropriate modifications to the frame design are necessary when using online suppliers. For example, the bottom edge of the glass can rest on mounting pads, and further reinforcement can be added to prevent deflection that would otherwise result from the less-distributed load.

[11] The words "Lite" and "Ultra-Lite" are used by 80/20 and TNutz to designate profiles that use less aluminum – and hence cost less – than their corresponding "standard" profiles.

Commercially available lightboards in 2025 are superior to and less expensive than those available in 2017. We are not comfortable recommending particular makes or models without testing or using them head to head. However, generally speaking, high quality rolling lightboards – with included frame-mounted presenter lighting – featuring 5-ft.-wide low-iron glass panes are now available for less than $4000. The quality and value are impressive. While – judging from online pictures – the casters on most commercially-available boards seem a bit small for heavy use, these rolling lightboards are otherwise well-sized for rolling into faculty offices. While $4000 may seem like a lot of money, a $4000 lightboard, shared by two faculty members for a period of 5 years, would represent an institutional per-faculty per-year expenditure of only $400. Meanwhile, at institutions with construction-capable faculty, staff, or students, a $2500 aluminum-framed institutional DIY build shared by two faculty members for 5 years would cost only $250 per faculty member per year. These per-year-per-faculty expenses are well within the faculty-development support budgets of many if not most higher education institutions.

## Conclusion

We hope that our build description and reflections are useful to lightboard designers and builders. We also hope that this paper increases interest in and support for office-based and other forms of personal lightboarding. More generally, we hope to contribute to an ever-expanding big-tent lightboarding perspective that values all contributors working in all contexts.

## Acknowledgments

We thank Merrimack College for supporting this 2017 build through a Provost Innovation Fund grant, which covered all materials expenses and which supported the student collaborators who carried out the bulk of the construction and who internally archived essential in-build and post-build documentation, including but not limited to the table in Appendix F and many photographs used throughout the paper and the Appendix.

[5]  Originally at: https://wiki.duke.edu/display/LIG/Lightboard. Subsequently moved to: https://duke.atlassian.net/wiki/spaces/LIG/pages/22085707/Lightboard. The build documentation is no longer available at either of those addresses as of 9/2025. However, the Duke University build can be essentially recovered by adding 12 inches to each of the long transverse T-slotted pieces in our build, using a 6-foot by 4-foot locally sourced low-iron glass glass pane (with appropriately located holes drilled prior to tempering), and by using the machine printed LED clips in [9] and either the original Duke standoffs [8] or the modified standoffs [7].

[6]  Craig W. Looney, 2018. https://youtu.be/i08xNFPBcwY

[7]  Nicholas A. Valente (2017 Merrimack standoffs).
     Sleeved part: https://www.thingiverse.com/thing:2329837
     Un-sleeved part: https://www.thingiverse.com/thing:2329168

[8]  Chip Bobbert (2014 Duke U. standoffs). https://www.thingiverse.com/thing:348064

[9]  Chip Bobbert (2014 Duke U. LED Clips). https://www.thingiverse.com/thing:379659

[10] Craig Looney. Making high quality videos on an inexpensive DIY lightboard (with technical tips relevant to all budgets). Spring 2023 Meeting of the New England Section of the American Physical Society, March 17-18, 2023, Amherst College, Amherst, MA. https://scholarworks.merrimack.edu/phy_facpub/20/

[11] Steve Griffiths, 2019. https://youtu.be/i68rwDF8ipA

[12] Craig Looney. Systematic Comparison of Blacklight vs. White Light for Lightboard Illumination. 2024 American Association of Physics Teachers Summer Meeting, July 6-10, 2024 (Boston, MA). https://scholarworks.merrimack.edu/phy_facpub/21/

[13] Sarah McCorkle and Paul Whitener (2017). The Lightboard: A faculty introduction to the development of supplemental learning media. 2017 Conference on Higher Education Pedagogy. Blacksburg, VA (February 2017). https://www.researchgate.net/publication/368471692_The_Lightboard_A_faculty_introduction_to_the_development_of_supplemental_learning_media

[14] Sarah McCorkle and Paul Whitener (2020). The Lightboard: Experiences and Expectations. *International Journal of Designs for Learning, 11*(1), 75–84. https://doi.org/10.14434/ijdl.v11i1.24642

# APPENDIX F: Frame Details

| Part Name | 80/20 Part # | Quantity | Label |
|---|---|---|---|
| 24" T-Slotted Profile | 1515 | 2 | A |
| 72" T-Slotted Profile | 1515 | 2 | B |
| 60" T-Slotted Profile | 1515 | 4 | C |
| 1515 45 Degree Support, 18" Long | 2545 | 2 pieces | D |
| 1515 45 Degree Support, 12" Long | 2535 | 6 pieces | E |
| 15 Series Gusseted Inside Corner Bracket | 4332 | 6 | F |
| 15 Series Tall Gusseted Inside Corner Bracket | 4336 | 2 | G |
| 15 Series 7-Hole Tee Flat Plate | 4312 | 4 | H |
| 15 Series 5 Hole "L" Flat Plate | 4481 | 2 | J |
| 15 Series Flange Mount Caster Base Plate | 2426 | 4 | K |
| Deluxe Flange Mount Swivel Caster | 2333 | 4 | L |
| T Handle Ball End Hex Wrench | 6000 | 3 | (tool) |
| 5/16-18 x .687" Flanged Button Head Socket Cap Screw (FBHSCS) | 3330 | 40 | {a} |
| Double economy T-Nut with 2 screws | 3355 | 26 | {b} |
| Triple economy T-Nut with 3 screws | 3357 | 6 | {c} |
| Roll-In T-Nut with Ball Spring | 13040 | 36 | {d} |

Capital letters A, B, … L correspond to labels in the pictures below. Note that the lightboard has left-right symmetry.

Bracketed lower case letters {a} … {d} are nuts and screws and are not shown in the figures below. Note that {b} and {c} are "economy" nut/screw combinations; the economy nuts must be slid into the slots prior to assembly. Meanwhile the roll-in nuts {d} can be "rolled in" to slots after assembly. They are more expensive but they provide more build-order flexibility.

All parts noted above are for 15 series T-slotted profiles; the part numbers are those for 80/20. Similar parts, with reasonably similar names, are available from Tnutz, normally at lower cost.

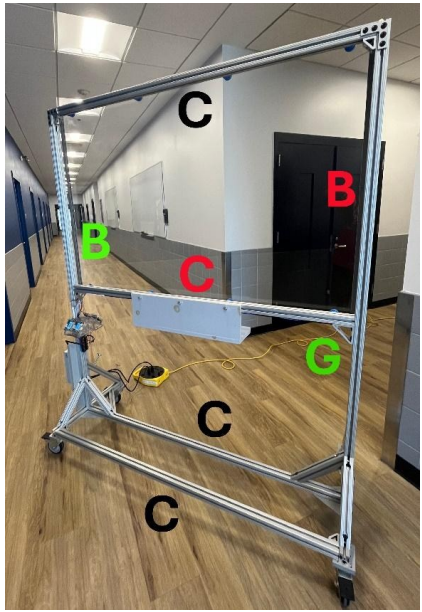
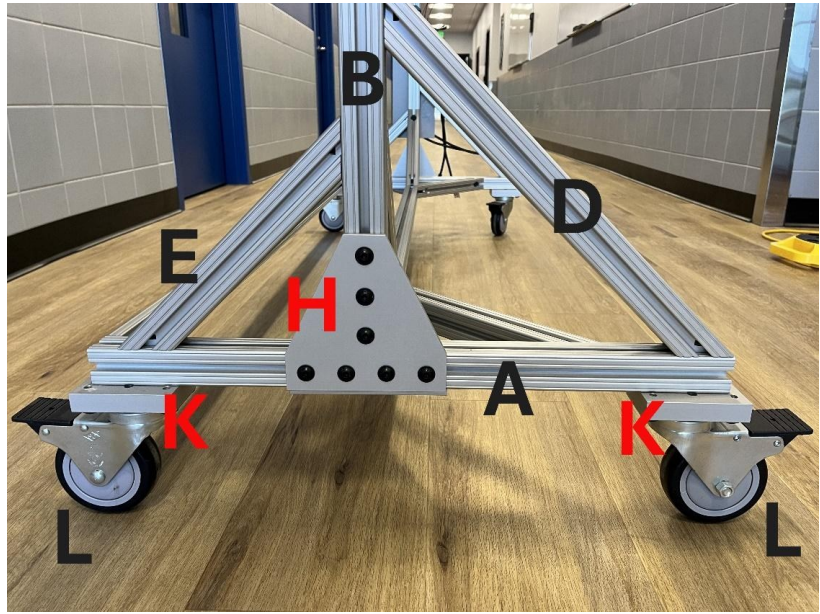

Note: the vertical supports are positioned so that the glass pane (which is substantially more massive than the vertical support members etc.) is centered within the base of support.

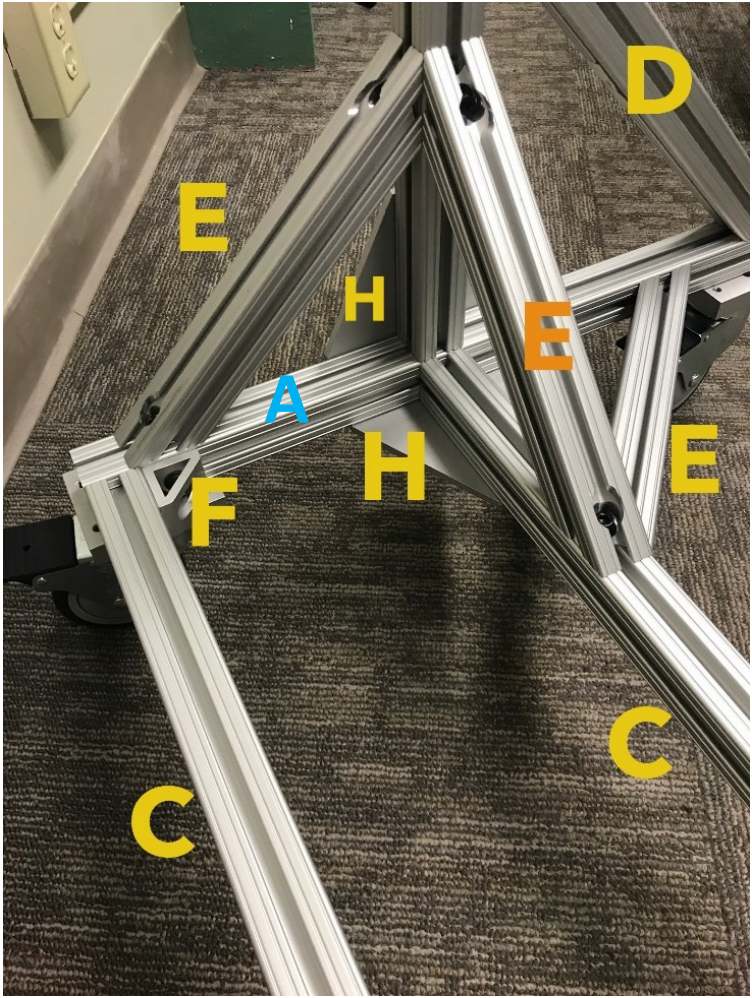
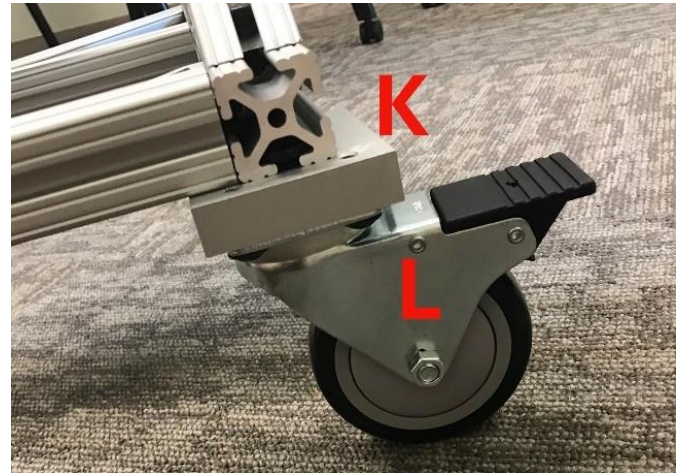
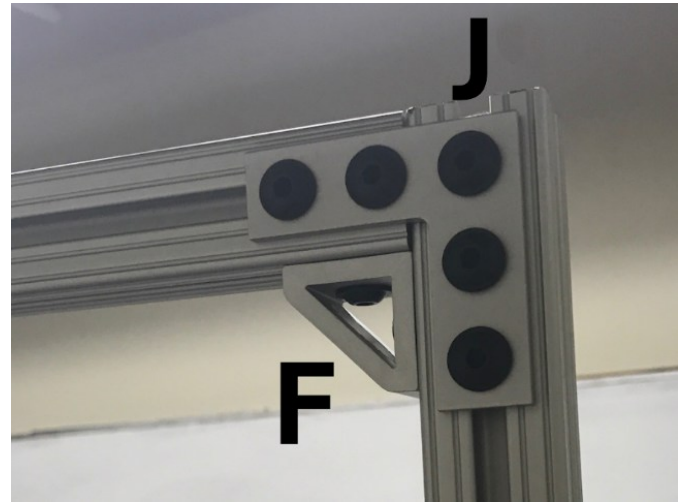